\documentclass[letterpaper,twocolumn,prl,showpacs,groupedaddress]{revtex4-1}%

\usepackage{amsmath}
\usepackage{graphicx}
\usepackage{dcolumn}
\usepackage{amssymb,amsmath}
\usepackage{color}
\usepackage{natbib}
\usepackage{amsfonts}
\usepackage{amssymb}%
\newcommand{\ba}{\begin{array}}
\newcommand{\ea}{\end{array}}
\newcommand{\be}{\begin{equation}}
\newcommand{\ee}{\end{equation}}
\newcommand{\bea}{\begin{eqnarray}}
\newcommand{\eea}{\end{eqnarray}}

\begin{document}

\title{Sedimentation of granular columns in the viscous and weakly inertial regimes}
\author{Hamza Chra\"ibi*$^{1}$ and Yacine Amarouchene*$^{1}$}
\affiliation{1 : Univ. Bordeaux, LOMA, UMR 5798, F-33400 Talence, France.\\
CNRS, LOMA, UMR 5798, F-33400 Talence, France.\\}
\date{\today }

\begin{abstract}
We investigate the dynamics of granular columns of point particles that
interact via long-ranged hydrodynamic interactions and that fall under the
action of gravity. We investigate the influence of inertia using the Green function
 for the Oseen Equation. The initial conditions (density and aspect ratio) are systematically
varied. Our results suggest that universal self similar laws may be sufficient
to characterize the temporal and structural evolution of the granular columns.
A characteristic time above which an instability is triggered (that may enable
the formation of clusters) is also retrieved and discussed.

\end{abstract}
\maketitle

Granular materials that are assemblies of discrete macroscopic solid particles
with sizes large enough that Brownian motion is irrelevant, have been a subject
of intensive research during the past few year \cite{Andreotti}. They are
ubiquitous in our everyday lives and remain at the heart of several
geophysical (sand dunes, coastal geomorphology, avalanches...) and industrial
processes (chemical, pharmaceutical, food, agricultural...) \cite{Andreotti}.
The variety of these fields make these granular materials subject to very
different flow and stress conditions. In particular, when the particles are
suspended in a fluid, one may expect that subtle hydrodynamic effects should
play a leading role \cite{Guazelli}. This must be contrasted with the case of
dry granular materials for which the influence of the carrying fluid is
negligible. In that case both the inelasticity of the collisions and/or the
friction between the grains are crucial \cite{Andreotti}.
While the falling of a single or couple of particles in purely viscous and
weakly inertial regimes was well described by Stokes and Oseen \cite{Brenner},
understanding the interactions of a cloud of particles remains a challenge, as
complex collective dynamics emerge due to the multiple long ranged
interactions (see fluidized beds \cite{Caflisch,GuazelliHinch}). Similar
difficulties exist also for n-body gravitational problems. Therefore, many
investigations were led in order to better understand the behavior of these
particle laden flows, presenting a large panel of geometries like jets,
streams, drops, spherical clouds. The sedimentation of spherical clouds of
particles, in an external fluid of variable viscosity, has been recently
investigated experimentally and numerically \cite{pignatel11}. At the
exception of the experimental work of Nicolas \cite{nicolas02}, investigations
related to jets or column of particles focused mainly on highly viscous fluids
(i.e. zero Reynolds numbers limit) \cite{pignatel09, crosby12}, air and
moderate vacuum (Large Reynolds numbers limit)
\cite{amarouchene2008,prado11,prado13} or other kinds of interactions :
\ capillary bridges, Wan Der Waals forces \cite{mobius06, royer09,
waitukaitis11, ulrich12}... ~~\\ \newline In this letter, we present an
investigation that fully characterizes, using point-particle simulations, the
dynamics of freely falling granular columns in different flow regimes,
clarifying the dependence to the Reynolds number, the aspect ratio and the
particle density.\\
The main characteristics of the
present system are: (i) solid particles suspended in a viscous fluid, and interacting by
virtue of the fluid, (ii) particles heavier than the fluid, thus sedimenting on account of
gravity. (iii) No continuous supply of particles in the granular cylinder.\\
 We will first describe the model used for the numerical
simulations, defining the characteristic quantities of the problem and its
relevant parameters before presenting and discussing our results.\newline At
the beginning of the simulation, we randomly initialize the positions of
$N_{0}$ particles in a cylindrical column of radius $R_{0}$ and length $H_{0}%
$, such as the dimensionless particle density $n_{0}=N_{0}/(\pi h^{\ast})$ is
homogeneous ($h^{\ast}=H_{0}/R_{0}$). In addition to their settling velocity
$U_{\eta}=F/(6\pi\eta a)$ in the fluid of viscosity $\eta$ under the action of
the gravitational force $F$, the point particles of radius $a$ are subject to
the hydrodynamic pairwise interactions modeled by the dimensionless Green function of the Oseen Equation
\cite{Brenner,Guazelli,pignatel11} which represents the additional velocity induced on a point particle by
another point particle distant by $\mathbf{d}=(d_{x},d_{y},d_{z})$ :
\begin{equation}
u_{k}^{\ast}=\frac{3}{4}a^{\ast}\left(  \frac{d_{k}}{d^{2}}\left[  \frac{2l^{\ast}}%
{d}(1-E)-E\right]  +\frac{E}{d}\delta_{kz}\right)  ~;~k=x,y,z\label{oseen}%
\end{equation}%
\begin{equation}
E=\exp\left(  -(1+\frac{d_{z}}{d})\frac{d}{2l^{\ast}}\right)  ~;~a^{\ast
}=a/R_{0}~;~l^{\ast}=\eta/(U_{\eta}\rho_{f}R_{0})
\end{equation}
In equation (\ref{oseen}), all lengths and velocities were made dimensionless
using $U_{\eta}$ as a reference velocity and $R_{0}$ as a reference length. A
reference time $\tau_{\eta}=R_{0}/U_{\eta}$ was also defined. $\rho_{f}$ is
the mass density of the external fluid and $l^{\ast}$ represents the
importance of the viscous effects. Note that the velocity given by equation (\ref{oseen}) is solution to the
corrected Navier-Stokes Equation, which models the weakly
inertial regime:
\begin{equation}
\rho_{f}(\mathbf{U_{\eta}}\cdot\mathbf{grad})\mathbf{u}=-\mathbf{grad}%
p+\eta\triangle\mathbf{u}~~~
\end{equation}
where $p$ is the fluid pressure.\newline By choosing the frame of reference
moving with the terminal settling velocity of an isolated particle, we compute
all the $N_{0}-1$ interactions on each particle and obtain a set of equations
describing the motions of the particles, of the form :
\begin{equation}
\frac{dM_{ki}}{dt}=\sum_{j\neq i}u_{k}^{\ast}~~~~~k=x,y,z~~i=1,N_{0}%
~~j=1,N_{0}%
\end{equation}
This equation is integrated using an Adams-Bashford time-marching algorithm
and at each iteration, we obtain the Cartesian position $M=(x,y,z)$ of each
particle. The detection of the interface of a granular column is performed by
dividing axially the domain into $h^{\ast}$ overlapping volumes. For each
volume, the radial position of the interface is calculated by calculating the
mean radial position of the farthest particles. The parameters for the
simulations are the aspect ratio $h^{\ast}$, the particle density $n_{0}$,
$l^{\ast}$ and $a^{\ast}$. However, one can notice that equation (\ref{oseen})
is linear with respect to $a^{\ast}$ and therefore the dynamics of the problem
will vary linearly with it. As a consequence, we set $a^{\ast}=0.05$ for all
simulations.
 As our simulations neglect particle-particle collisions 
they are only applicable to dilute regimes. \newline It is interesting to write the particle Reynolds
number such as $Re_{p}=a^{\ast}/l^{\ast}=aU_{\eta}\rho_{f}/\eta$ which
variation in this problem is performed by varying $\eta$, however as we are
investigating the behavior of a macroscopic object, we have to define a
macroscopic Reynolds number $Re=R_{0}U_{col}\rho_{f}/\eta$, where $U_{col}$ is
the characteristic velocity of a cylindrical column. $U_{col}$ can be defined
using the settling velocity of a vertical cylinder of aspect ratio $h^{\ast}$
in a viscous fluid, hence $U_{col}=P\ln(h^{\ast}/2)/(2\pi\eta H_{0})$, $P$
being the macroscopic gravitational force. Calculating the equivalent mass of
the column from its volume fraction, one can find that $U_{col}=3\pi
n_{0}a^{\ast}\ln(h^{\ast}/2)U_{\eta}$ and therefore $Re=3\pi n_{0}\ln(h^{\ast
}/2)Re_{p}$.\newline
\begin{figure}[h]
\begin{center}
\includegraphics[width=1.\columnwidth]{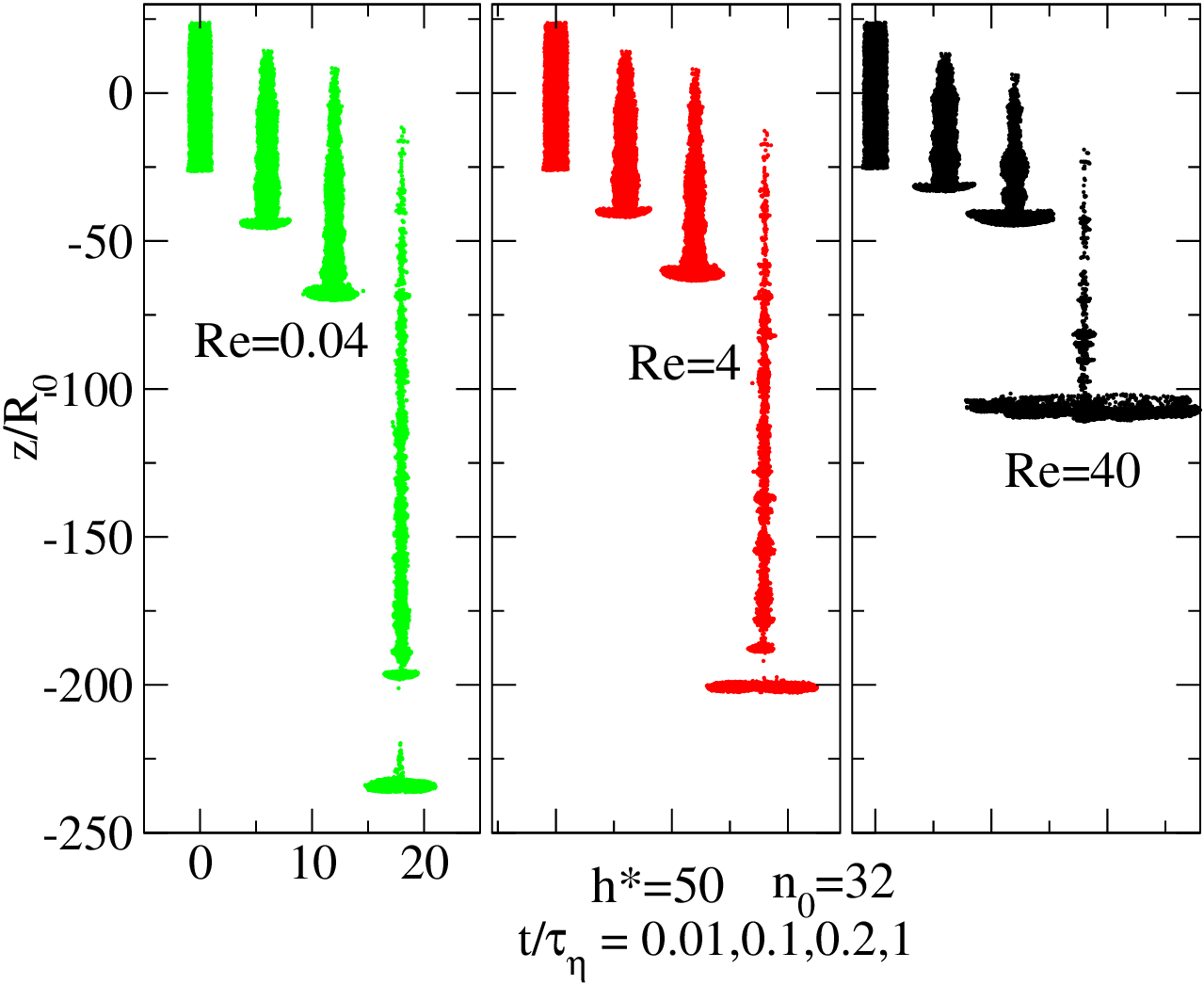}\label{columns}
\end{center}
\caption{(Color
  online)Falling of cylindrical granular columns for $Re=0.04$ , $4$ and $40$
with an aspect ratio $h^{\ast}=50$ and particle density $n_{0}=32$. Four
different instant are shown such as $t/\tau_{\eta}=0.01$, $0.1$, $0.2$ and
$1$, and time increases from left to right. The columns are shown in the
reference frame of an isolated particle falling at its settling velocity.}%
\end{figure}
~~\\
An example of simulations performed varying $Re$ are shown in figure 1. We can
first observe that while they fall, all the columns stretch and thin. We can
also observe that a leading mushroom shaped plume forms at the front \cite{pignatel09} while a
particle leakage can be observed at the rear. For the last instant, we can see
that the columns lose their cohesion and eventually detach into shorter
columns and droplets, which means that a varicose instability grows in time.
Considering now the effect of the Reynolds number in relative frames, we can
see that increasing $Re$ relatively slows down the falling of the columns and
increases their effective cohesion. We can also observe that size of the
leading mushroom is larger for a same instant. It is important to understand
that, in order to compare them, columns for different $Re$ were represented in
different frames. As the viscosity of the fluid for $Re=0.05$ is much larger
than the viscosity for $Re=50$, columns at large $Re$ in the absolute frame
will experience a faster dynamics. These results are qualitatively comparable
to those of Pignatel \cite{pignatel11}, which showed that increasing $Re$
enhances effective cohesion and slows the falling of spherical clouds of
particles. \newline\begin{figure}[h]
\begin{center}
\includegraphics[width=1.\columnwidth]{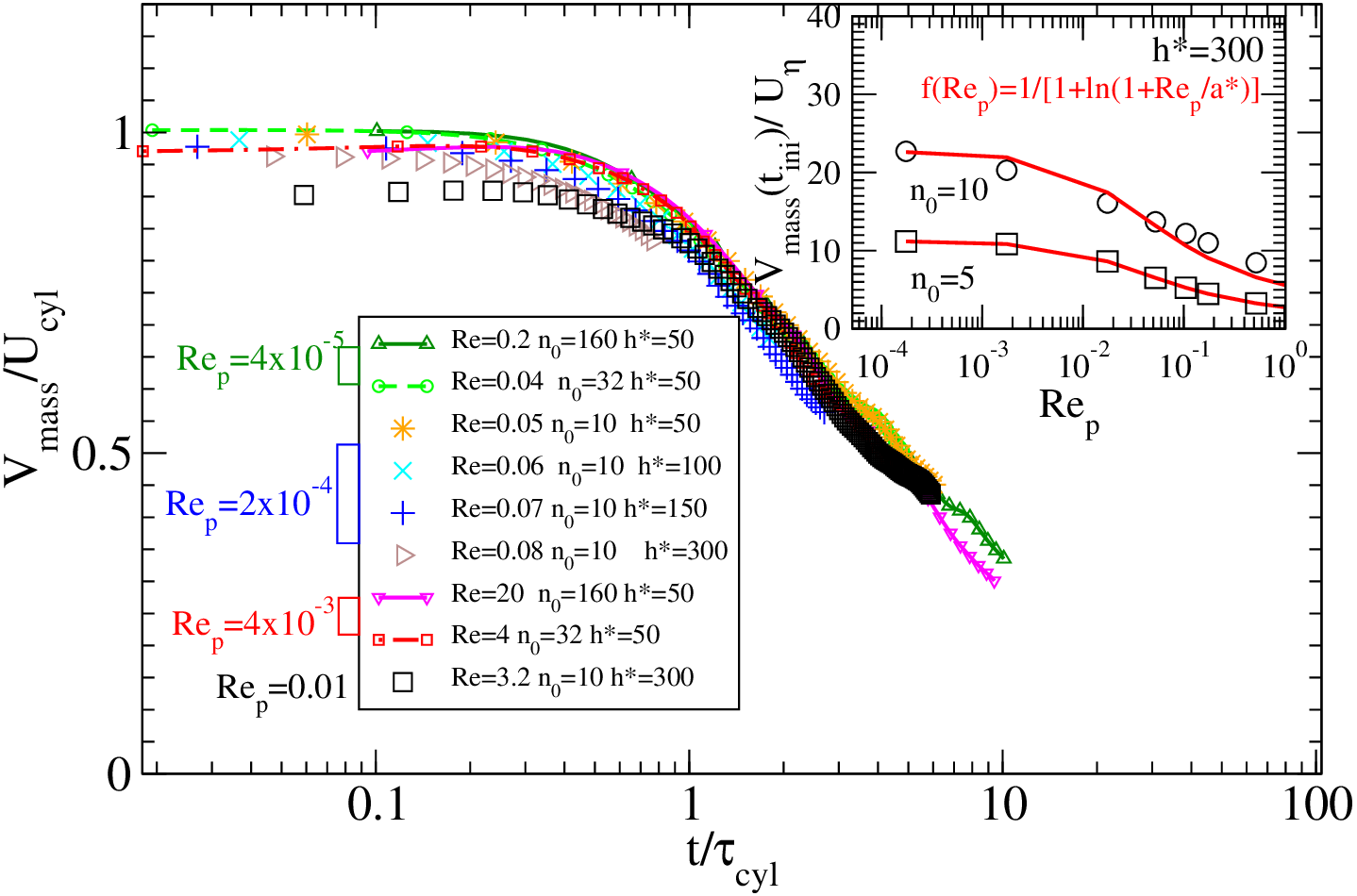}
\end{center}
\caption{(Color
  online)Variation of the reduced velocity of the center of mass
$V_{mass}/U_{cyl}$ versus reduced time $t/\tau_{cyl}$ for $Re_{p}%
=4\times10^{-5},2\times10^{-4},4\times10^{-3}$ and $10^{-2}$, $h^{\ast}=50,100,150,300$ and
$n_{0}=10,32,160$; $U_{cyl}=U_{col}/(1+\ln(1+Re_{p}/a^{\ast}))$ and
$\tau_{cyl}=H_{0}/U_{cyl}$. Inset: variation of dimensionless initial velocity
of the center of mass $V_{mass}/U_{\eta}$ versus $Re_{p}$ for $n_{0}=5$, $10$
and $h^{\ast}=300$ (symbols : simulations, line : analytical function).}%
\end{figure}
~~\\
Figure 2 shows the time variation of the center of mass $V_{mass}$ reduced by
a corrected characteristic velocity of the column $U_{cyl}$. Indeed, when
varying $Re_{p}$ at fixed $h^{\ast}$ and $n_{0}$, we observed a correction in
$V_{mass}$ which was not taken into account in $U_{col}$. This correction is
shown in the inset of figure 2, where $V_{mass}/U_{\eta}$ decreases with
$Re_{p}$ following a logarithmic behavior. We successfully retrieved this
behavior by the function $f(Re_{p})=1/(1+\ln(1+l^{\ast-1}))=1/(1+\ln
(1+Re_{p}/a^{\ast}))$ and in order to take into account the $Re_{p}$
dependence of the dynamics, we defined a new characteristic velocity
$U_{cyl}=3\pi U_{\eta}n_{0}a^{\ast}\ln(h^{\ast}/2)f(Re_{p})$ along with a
macroscopic characteristic time $\tau_{cyl}=H_{0}/U_{cyl}$. Finally, we can
observe in figure 2 that for a large set of different parameters, all the
temporal evolution of $V_{mass}/U_{cyl}$ collapse in a single universal curve.
It shows that for $t\ll\tau_{cyl}$, the column fall with a constant velocity
$U_{cyl}$ before decreasing with time following a logarithmic behavior when
$t\gg\tau_{cyl}$. This confirms that $U_{cyl}$ and $\tau_{cyl}$ are the
adequate velocity and characteristic time that describe the falling of the
columns.\newline\begin{figure}[h]
\begin{center}
\includegraphics[width=1.\columnwidth]{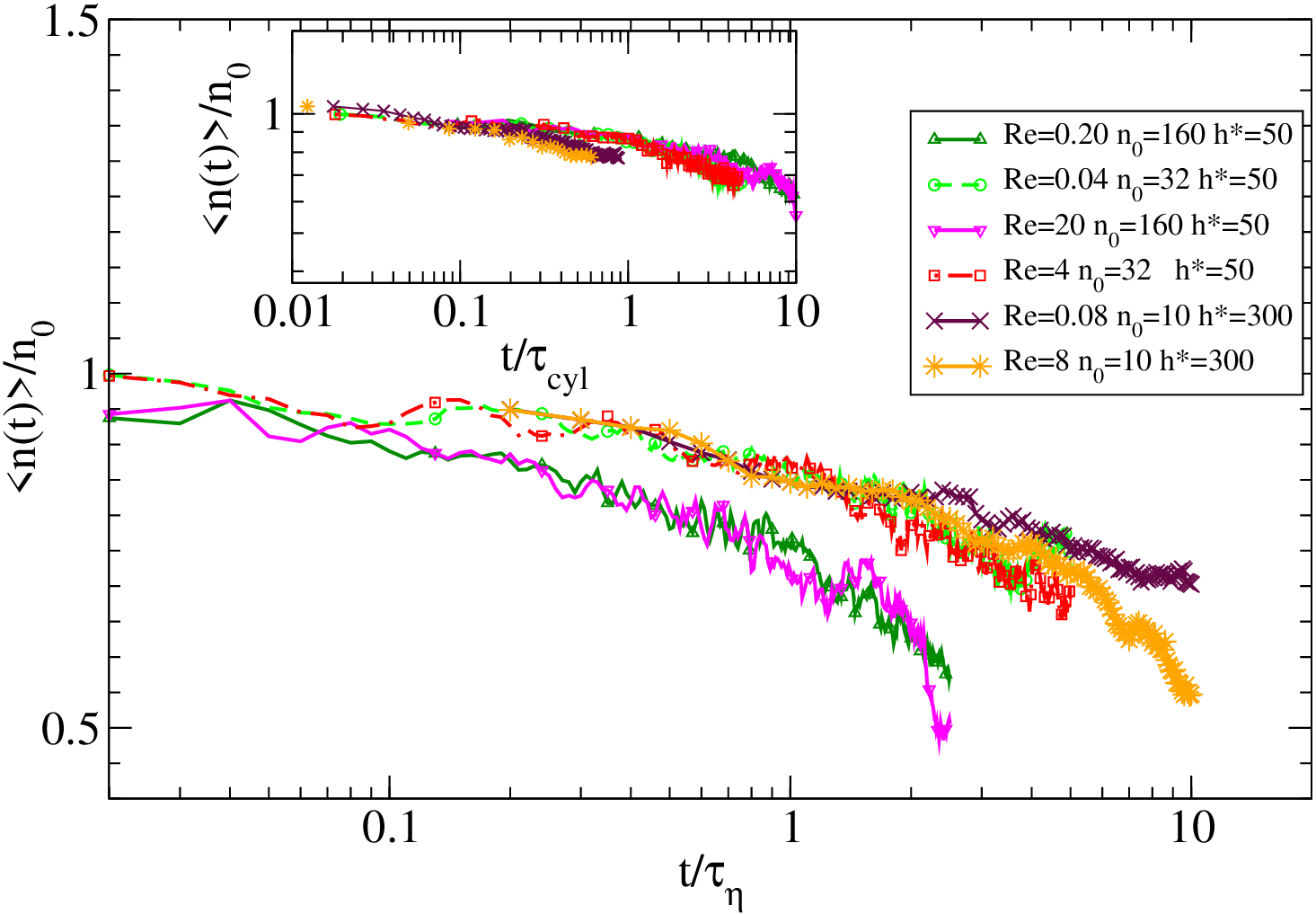}
\end{center}
\caption{(Color
  online)Variation of the reduced particle mean density $<n(t)>/n_{0}$ versus
dimensionless time $t/\tau_{\eta}$ for $Re_{p}=4\times10^{-5}$, $4\times10^{-3}$ and
$n_{0}=32$ and $160$. Inset : Variation of the reduced particle mean density
$<n(t)>/n_{0}$ versus the reduced time $t/\tau_{cyl}$ for the same
parameters.}%
\end{figure}
~~\\
The variation of particle mean density $<n(t)>/n_{0}$ versus
dimensionless time $t/\tau_{\eta}$ is presented in figure 3. For different
Reynolds number and initial particle densities, we observe a first
incompressible regime where $<n(t)>$ is almost constant followed by a weakly
compressible regime where the mean particle density experiences a slow time
decay. While in the main figure, there seems to be different dynamics, the
inset of figure 3 shows that using the characteristic time $\tau_{cyl}$
provides a better collapse of the data. In addition, we can see that $t\ll
\tau_{cyl}$ corresponds to an incompressible regime while $t\gg\tau_{cyl}$
corresponds to a weakly compressible flow.
\bigskip\begin{figure}[ptb]
\begin{center}
\includegraphics[width=1.\columnwidth]{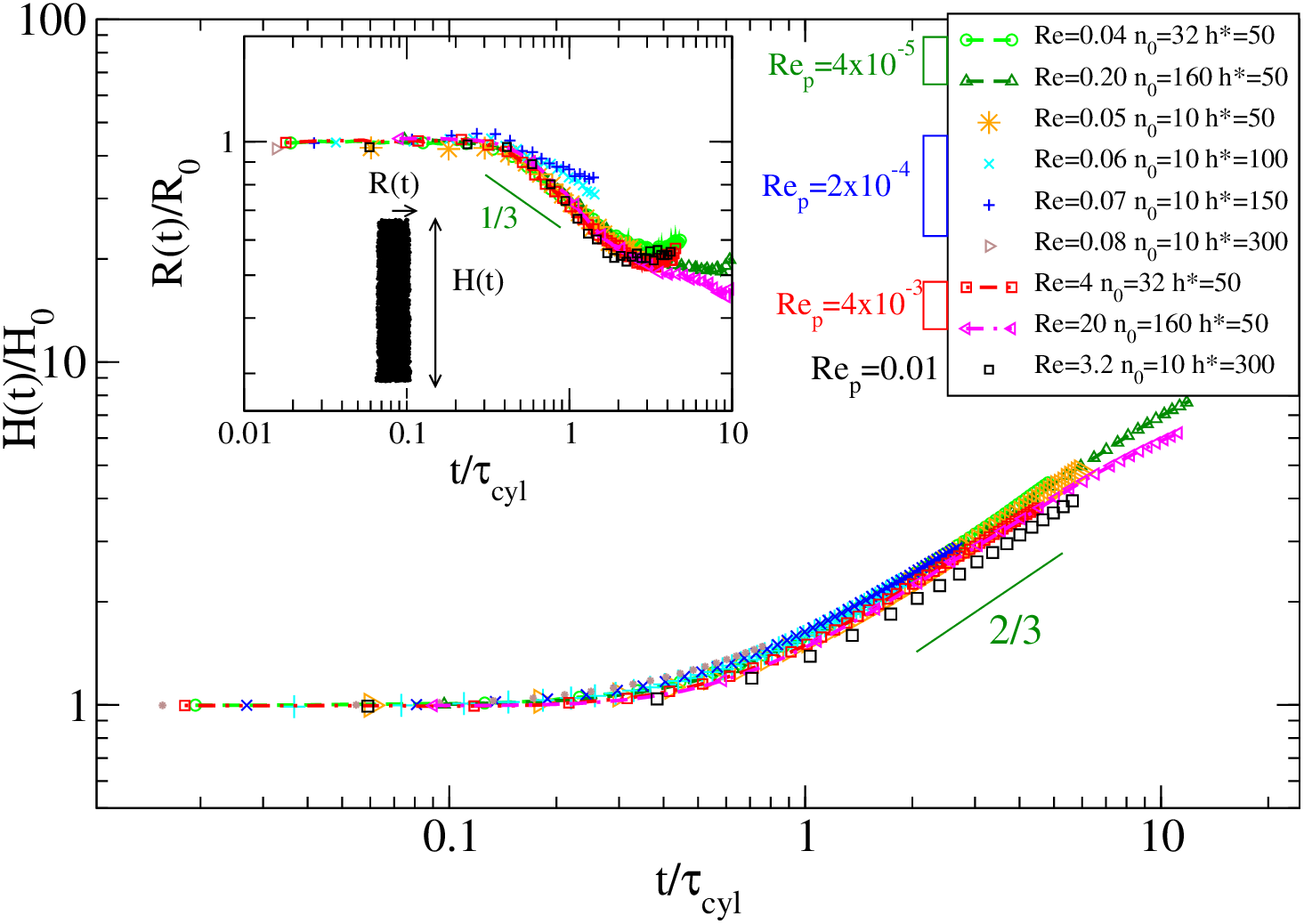}
\end{center}
\caption{(Color
  online)Variation of the reduced length of the column $H(t)/H_{0}$ versus
reduced time $t/\tau_{cyl}$ for $Re_{p}=4\times10^{-5},2\times10^{-4},4\times10^{-3}$ and
$10^{-2}$, $h^{\ast}=50,100,150,300$ and $n_{0}=10,32,160$. Inset: Variation
of the mean reduced radius of the column $R(t)/R_{0}$ versus reduced time
$t/\tau_{cyl}$ for the same set of parameters.}%
\end{figure}
~~\\
The deformations of the columns are displayed in figure 4. It provides an
adequate description of the dynamics of both the reduced length $H(t)/H_{0}$
and of the reduced mean radius $R(t)/R_{0}$ (calculated excluding the
extremities of the column). Once again, the dynamics of the column deformation
for a large set of different parameters $h^{\ast}$, $n_{0}$ and $Re_{p}$ are
represented by universal curves. When $t\ll\tau_{cyl}$ the columns remain
undeformed while for $\tau_{cyl}<t<10\tau_{cyl}$, the length increases
following a universal scaling such as $H(t)\sim H_{0}(t/\tau_{cyl})^{2/3}$ and
the mean radius decreases such as $R(t)\sim R_{0}(t/\tau_{cyl})^{-1/3}$.
Assuming a weakly compressible flow for $t>\tau_{cyl}$ (in agreement with
figure 3), the volume of the column has to remain constant, i.e. $\pi
R(t)^{2}H(t)\sim\pi R_{0}^{2}H_{0}$ which is well recovered by the previous
scaling laws.
\newline Now let us focus on the strain rate
$dV_{z}/dz$ selected at a local scale. It is an important parameter to
describe the elongational stretching applied to the columns. It is known that
stretching stabilizes liquid columns and prevent instabilities from growing
and forming satellite drops \cite{eggers97}. Figure 5 provides a universal
curve representing the variation of the reduced axial velocity gradient
$(dV_{z}/dz)/(n^{\ast}U_{\eta}/H_{0})$ versus reduced time $t/\tau_{cyl}$
where different set of parameters present a good collapse. The elongation
rate, or axial velocity gradient (which extent grows with time along the column),
 is deduced from the axial velocities of the
particles at the rear of the columns, as shown in the inset. We can clearly
see that in the incompressible regime ($t<\tau_{cyl}$), the elongational rate
remains constant and scales like $n^{\ast}U_{\eta}/H_{0}$. This scaling comes
from the fact that a single particle is on average surrounded by
$2N_{0}/h^{\ast}=n^{\ast}$ particles (i.e. the particles contained in a sphere
of diameter $2R_{0}$), therefore its characteristic velocity is $n^{\ast
}U_{\eta}$ while its characteristic axial length is $H_{0}$. In the weakly
compressible regime ($t>\tau_{cyl}$), $dV_{z}/dz$ decays like $t^{-1}$. This
is consistent with an incompressible self similar decay of the column radius
$R(t)\sim R_0 t^{\alpha}$ that gives $(dV_{z}/dz)=-\frac{2}{R}\frac{dR}{dt}\sim
t^{-1}$ independently of the thinning exponent \ $\alpha$. Although not
strictly comparable as they perform event-driven simulations of streams of
particles interacting via collisions and cohesive forces and not via
hydrodynamic interactions, Ulrich and Zippelius \cite{ulrich12} showed a
similar result in the case of particles that fall in vacuum under the action
of gravity. In that case the elongational rate is simply retrieved from the
incompressibility condition and the velocity field imposed by the free fall.
Finally, figure 5 also displays snapshots of columns at different Reynolds
number. We observe that for $t\ll\tau_{cyl}$, the column are cohesive and
destabilization has not yet occurred, while the columns for $t>\tau_{cyl}$ show
a clear destabilization due to the development of a varicose instability.
\newline\begin{figure}[h]
\begin{center}
\includegraphics[width=1.\columnwidth]{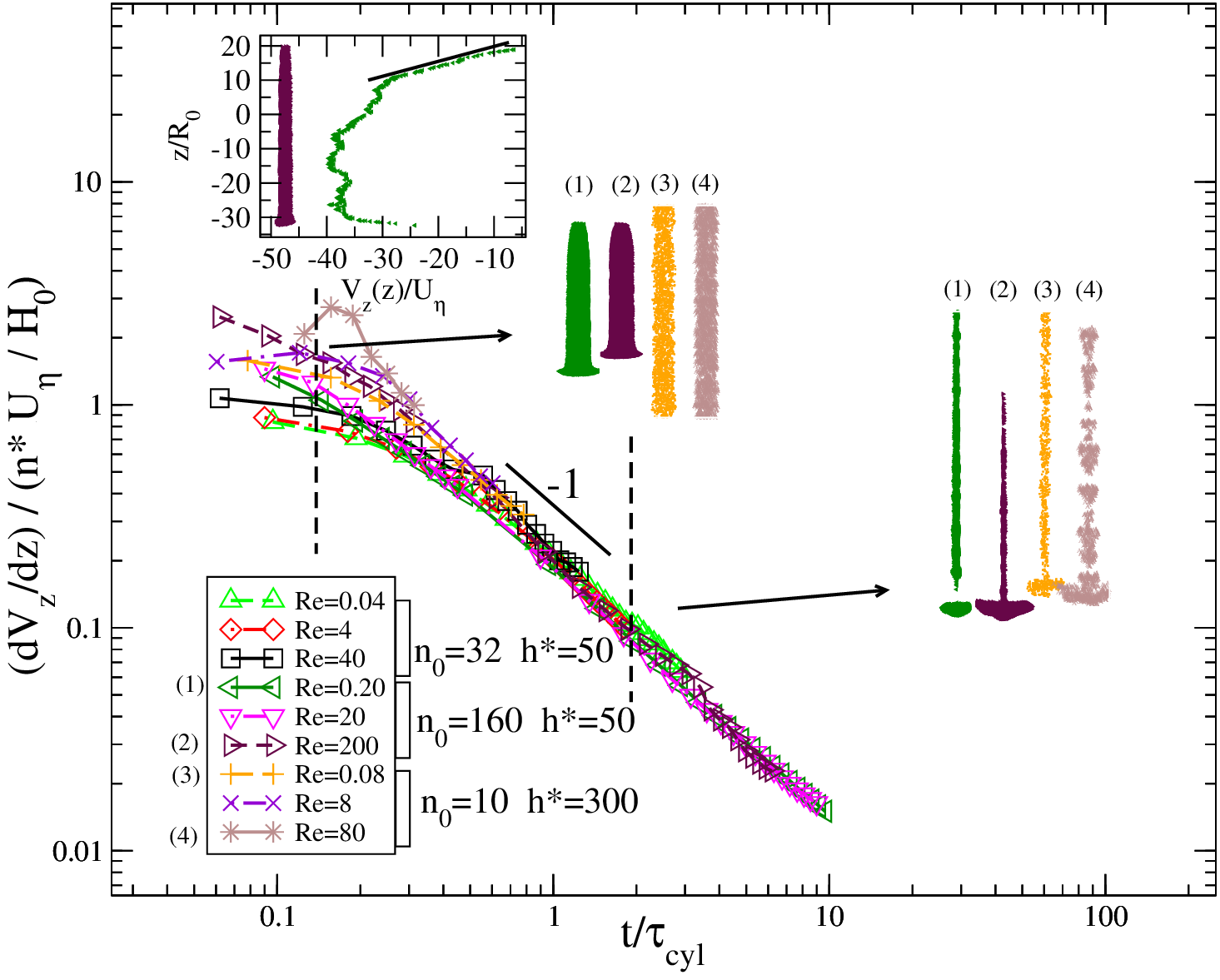}
\end{center}
\caption{(Color
  online)Variation of the reduced axial velocity gradient $(dV_{z}%
/dz)/(n^{\ast}U_{{}}\eta/H_{0})$ versus reduced time $t/\tau_{cyl}$ for
$Re=0.04....200$, $h^{\ast}=50,300$ and $n_{0}=10,32,160$; $n^{\ast}=2\pi
n_{0}$. Inset: Shape of a column (left) axial position $z/R_{0}$ versus
particle axial velocity $V_{z}$ (right). The black line, shows the linear
behavior of $V_{z}$ with $z$.}%
\end{figure}
~~\\
Finally, let us focus on the description of the instability that leads to the
destabilization of the columns. The main features of the instability are shown
in figure 6. In the main panel of figure 6, we can note that the value of the
most unstable wavelength $\lambda$ is almost constant and shows no clear
dependence on the Reynolds number ($\lambda$ was deduced from the interface
profile). In the regime $Re<<1$, we found $\lambda
\sim15R_{0}$ for $n_{0}=5$ and $\lambda\sim12R_{0}$ for $n_{0}=10$ which are
both consistent with the values found in earlier investigations
\cite{pignatel09,crosby12} dedicated to the effect of $n_{0}$. \ The inset of
figure 6 shows the temporal evolution of the standard deviation of the reduced
radial variation (excluding the front drop) $\sigma n_{0}^{1/2}$ for $n_{0}=5$
and $10$ and for $Re=0.08, 0.8, 8$.
We can see that the data collapse for $Re\ll1$, in good agreement with Crosby
and Lister who suggested that the growth of the standard deviation of the
reduced radial variation are mainly due to fluctuations in the average number
density of particles along the axial distance about its mean value
\cite{crosby12} . However, $\sigma$ seems to present larger values for
$Re\gg1$.
This means that increasing the Reynolds number may have a noticeable
effect on the varicose instability. This induces a stronger effective cohesion
and leads to a more efficient destabilization. These observations provide
another route to the instability of granular jets along with the recently
observed clustering due to cohesion and liquid bridges between grains
\cite{royer09,waitukaitis11}.
 Furthermore, our results suggest that the sedimentation of
particle-laden jets may eventually furnish an interesting system to study the
compressible Rayleigh -Plateau instability as suggested recently
\cite{Miyamoto}. \\
\begin{figure}[h]
\begin{center}
\includegraphics[width=1.\columnwidth]{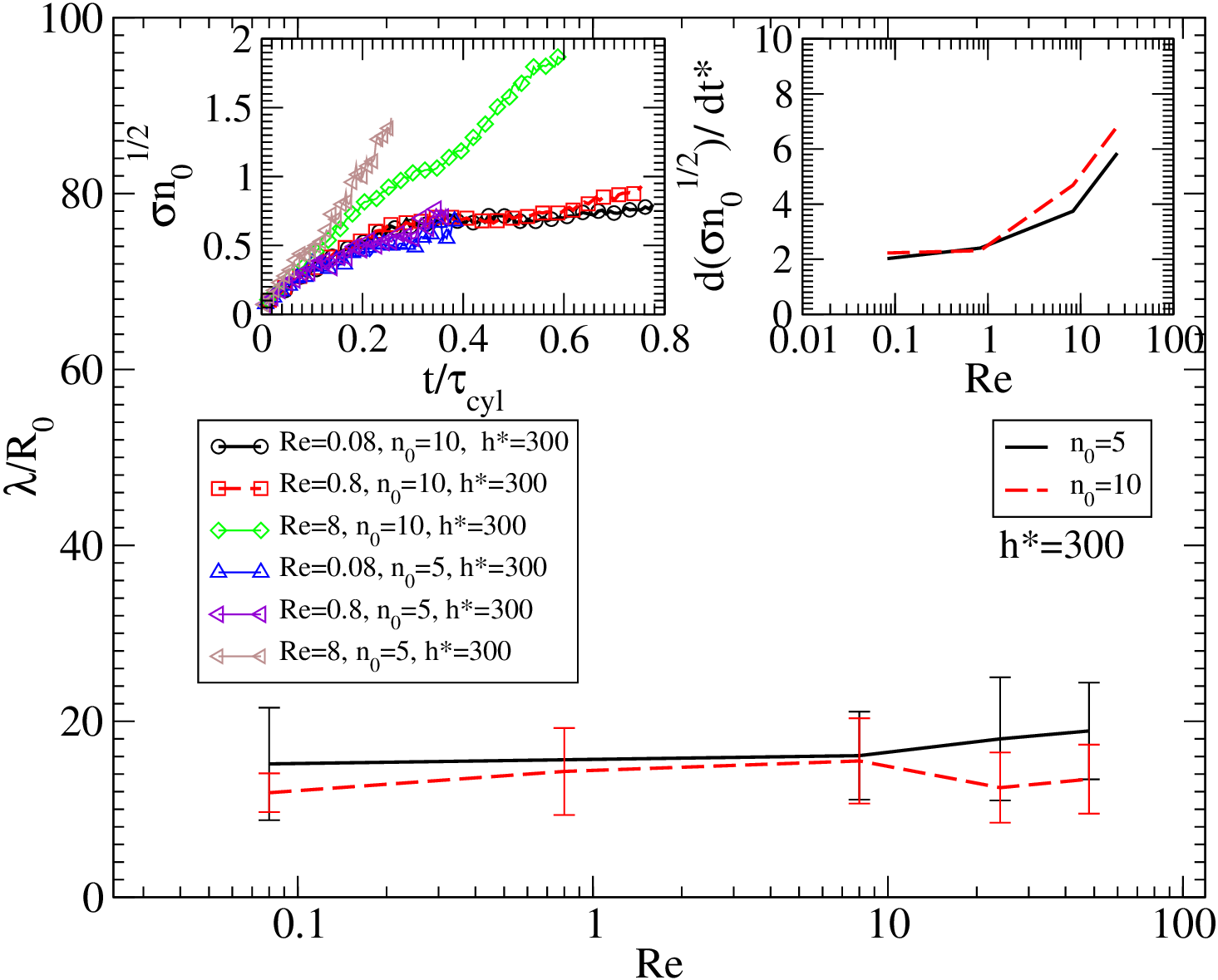}
\end{center}
\caption{(Color
  online)Variation of the reduced most unstable wavelength $\lambda/R_{0}$
versus $Re$ for $n_{0}=5,10$ and $h^{\ast}=300$. Left inset : Variation of the
standard deviation of the radial variation $\sigma$ versus reduced time
$t/\tau_{cyl}$ for $n_{0}=5,10$, $Re=0.08,0.8,8$ and $h^{\ast}=300$. Right inset : Variation 
of $d (\sigma n_0^{1/2}) / dt^*$ (calculated at short times) versus $Re$ for $n_{0}=5,10$ and $h^{\ast}=300$. $t^*=t/\tau_{cyl}$.}%
\end{figure}
~~\\
\newpage
To conclude, we have shown that universal scaling laws fully characterize the
dynamics of free falling granular columns in viscous fluids. The
characteristic velocity $U_{cyl}$ scales linearly with the particle density,
while it shows a logarithmic increase with the aspect ratio and a decreasing
logarithmic correction with the particle Reynolds number. A universal
characteristic time $\tau_{cyl}$ based on $U_{cyl}$ and the column length
$H_{0}$ has also been retrieved. When $t<\tau_{cyl}$, the flow could be
considered as incompressible, and the columns deform only slightly and are
subjected to a constant strain rate $n^{\ast}U_{\eta}/H_{0}$ while falling at
a constant velocity $U_{cyl}$. For $t>\tau_{cyl}$, we showed that the flow was
weakly compressible, and that the columns were subjected to an elongational
rate decaying like $t^{-1}$, while they stretched like $t^{2/3}$ and thinned
like $t^{-1/3}$ before the development of a varicose instability leading to a
long wavelength destabilization.
Finally, we found that the most unstable wavelength of the instability of the
order of $\sim10R_{0}$ is almost independent of inertia corrections while the
growth rate of the most unstable mode shows a clear increase with the Reynolds number.\\
{\bf Acknowledgement}\\
We thank Pierre Navarot for the  preliminary investigation . This research is
supported by CR Aquitaine Grants no. 2006111101035, no. 20091101004 and by ANR
Grant no. ANR-09-JCJC-0092.\\
{\small * Corresponding authors' e-mails :\\ h.chraibi@loma.u-bordeaux1.fr\\
y.amarouchene@loma.u-bordeaux1.fr} \\

\end{document}